# Towards Zero Trust Architecture：A Pilot Study on Information Systems Security Readiness amongst Small and Medium Enterprises

*Completed Research Paper*

**Yu Deng**
Department of Accounting and Information Systems,
University of Canterbury,
New Zealand

**Anushia Inthiran**
Department of Accounting and Information Systems,
University of Canterbury,
New Zealand

## Abstract

*Small and medium enterprises (SMEs) face growing cyber threats but often lack the resources and expertise needed to adopt Zero Trust Architecture (ZTA). This pilot study examines the drivers and barriers shaping SME perceptions of ZTA necessity and proposes an exploratory staged adoption path. Survey data from 64 IT and security professionals in the Asia-Pacific region show that ZTA familiarity and cloud-computing needs are the strongest positive correlates of perceived necessity, whereas accumulated barriers show only a weak negative association. Identity and access management complexity and scalability emerge as the main implementation hurdles. Based on these findings, we propose a three-stage route for SMEs: strengthening identity governance, segmenting high-value assets, and introducing targeted monitoring in line with operational capacity. The study offers early evidence for more realistic Zero Trust transitions in resource-constrained firms.*

**Keywords:** Zero trust architecture (ZTA), small and medium enterprises (SMEs), security readiness, pilot study, cyberthreats

## Introduction

Traditional network security architectures often adopt the design paradigm of “strong on the outside, weak on the inside”, creating a vulnerable “fragile intranet” structure that makes boundary-based protection models susceptible to penetration by skilled attackers (Tyler & Viana, 2021). As small to medium enterprises (SMEs) embark on digitalization, cloud transformation initiatives and grapple with data storage issues, traditional security mechanisms based on network boundaries face significant limitations (He et al., 2022). The core shortcomings of traditional network security models are reflected in three areas: the assumption of excessive trust in internal traffic, the lack of fine-grained access control mechanisms, and the difficulty of adapting to new information technology practices such as cloud computing and telecommuting. Research has shown that even when deployed consistently as a foundational security component, traditional architectures are still ineffective against attacks from advanced persistent threats (APTs) (Kanei et al., 2021).

Zero Trust Architecture (ZTA) breaks away from the traditional paradigm of relying on firewalls, Intrusion Detection Systems (IDS) or Intrusion Prevention Systems (IPS) to construct network boundaries, and instead enforces dynamic security policies for each individual connection between users, devices, applications and data (Lindemulder & Kosinski, 2024). The core features of this architecture is manifested in the denial of a default state of trust to any user or device, regardless of whether it is inside or outside the organisational network (Rose et al., 2020), and the reconfiguration of the resource protection system through identity-centric and context-aware fine-grained control mechanisms (Hirai et al., 2023).

The motivation for this research arises from the increasing recognition of ZTA as a promising solution for enhancing organisational cybersecurity. However, despite its advantages, concerns persist regarding its practical implementation especially amongst SMEs. Overall, organisations that have implemented ZTA are primarily large enterprises or data security sensitive organisations, however the uptake of ZTA within SMEs is low (Grady, 2024). SMEs play a critical role in the global economy. According to the World Bank, SMEs represent over 90% of businesses and over 50% of employment worldwide (The World Bank, 2022). Consequently, SMEs have an extensive and pressing need for robust cybersecurity measures. However, despite SMEs significance, they have largely been overlooked in cybersecurity initiatives (Junior et al., 2023).

Existing work on SME cybersecurity highlights that smaller firms are "unaware, unfunded and uneducated" relative to their exposure, with persistent gaps in security capability, governance, and investment (Junior et al., 2023; Vocalcom, 2014). This structural position in the economy, combined with weaker security baselines, makes SMEs both attractive targets and vulnerable nodes in supply chains, where compromise can propagate to larger organisations. These features create a distinct context in which to study ZTA. Resource constraints, lean IT teams, and reliance on commodity cloud services mean that SMEs are unlikely to adopt the same architectures, migration paths, or governance arrangements as large enterprises. At the same time, regulatory and customer pressures increasingly extend to smaller suppliers, making auditable controls and traceable access more salient. Recent conceptual work argues that ZTA could serve as a risk countermeasure in SMEs and advanced-technology systems (Abdelmagid & Diaz, 2025), but offers limited empirical evidence on how SME practitioners actually perceive ZTA, which factors shape their sense of necessity and advantage, and how they envisage feasible adoption paths.

Studying ZTA from an SME perspective is therefore necessary for at least three reasons. First, it addresses an empirical blind spot in the security and IS adoption literature, where insights derived from large enterprises cannot be assumed to transfer to smaller firms. Second, it informs whether and how ZTA can be translated into staged, governable adoption paths that fit SME constraints rather than replicating large-enterprise blueprints. Third, it provides a basis for tailoring guidance, managed services, and policy instruments so that they support realistic Zero Trust practices in SMEs.

This study examines perceptions of, and governance for ZTA within the SME context. Although prior IS security adoption research has examined organisational capability, perceived risk, and protective motivation, less is known about how these factors shape SME perceptions of ZTA specifically. We formulate two research questions: RQ1: What are the levels of perceived necessity of ZTA and its relative competitive advantage? RQ2: What organisational and environmental factors, specifically IT capabilities and cloud dependencies, are associated with perceived necessity in this exploratory pilot sample? Using a sample of 64 practitioners and applying non-parametric tests, correlation analysis, and regression with ordered-logit robustness checks, we provide exploratory evidence on ZTA-related perceptions and develop practical implications for staged governance in SME contexts.

# Background

## *Zero Trust Architecture Overview*

ZTA rests on the principle of "never trust, always verify" and shifts defence from network perimeters to continuous, identity-centric controls (Rose et al., 2020). Authoritative guidance converges on five pillars—identity, devices, networks, applications & workloads, and data—supported by visibility & analytics, automation & orchestration, and governance as cross-cutting capabilities (Cybersecurity and Infrastructure Security Agency [CISA], 2023). At scale, ZTA has moved beyond concept to policy-mandated implementation: the U.S. Federal Zero Trust Strategy (OMB M-22-09) requires agencies to meet specific

ZTA objectives, while the U.S. Department of Defence issued a department-wide Zero Trust Strategy and capability roadmap (DoD, 2022; OMB, 2022). In industry, Google's BeyondCorp program operationalised ZTA principles enterprise-wide, documenting design, migration, and the 'long tail' of complex use cases. National cyber authorities in the UK, Australia, and New Zealand provide prescriptive ZTA guidance that organisations can adopt or tailor (ACSC, 2025; NCSC, 2022; NZISM, 2023). Reported benefits of having implemented ZTA include reduced lateral movement through granular policy enforcement, continuous authentication and authorisation across hybrid environments, and improved resilience via strong device posture and observability. Government strategies position ZTA as a means to meet measurable security objectives by deadlines, while enterprise case studies show that productivity can be maintained during migration when access is decoupled from network location and governed by risk (DoD, 2022; Gonçalves, 2023; NCSC, 2022; OMB, 2022). These policy and practitioner sources are used here to define the applied ZTA context and to summarise implementation guidance. The theoretical interpretation of ZTA readiness in this study is developed through IS security adoption research and the TOE and PMT lenses discussed below.

## *Cyber-security Realities of SMEs*

SMEs typically operate with constrained budgets and limited specialist staff, which weakens identity management and endpoint compliance and, in turn, elevates exposure to ransomware, business email compromise, and supply-chain compromise. A recent Australian report indicates SMEs persistently experience high volumes of cybercrime affecting small businesses, while European Union wide statistics show that only 53% of SMEs maintain a formal security policy compared with 85% of large enterprises (ACSC, 2025; ENISA, 2024). This discrepancy highlights the urgent need for effective, scalable security measures tailored to the resource constraints of SMEs.

Despite the growing recognition of ZTA as a robust security model, its adoption among SMEs has been slow. SMEs often rely on traditional perimeter-based security models that are ill-equipped to address modern cyber threats, such as APTs and insider attacks, which are increasingly bypassing these legacy defences (Kanei et al., 2021).

Currently, research into ZTA adoption in SMEs is scarce, and the barriers they face in its implementation remain underexplored. Existing studies tend to focus on large enterprises, where the technical infrastructure and budgetary support for ZTA are more readily available. In contrast, SMEs are often overlooked in cybersecurity initiatives, despite the fact that they face significant security risks, including exposure to ransomware, business email compromise, and supply-chain attacks.

Furthermore, regulatory and compliance pressures are increasing for SMEs, with many facing heightened scrutiny around data protection and privacy regulations. Implementing ZTA can help SMEs align with these requirements by enabling better traceability and auditability of access and usage patterns. This is particularly critical as the EU's General Data Protection Regulation (GDPR) and other data protection laws impose stringent requirements on organisations to protect personal data from unauthorized access.

From an IS adoption perspective, these constraints mean that ZTA readiness cannot be inferred solely from the availability of technical guidance; it also depends on organisational capability, perceived feasibility, and decision-makers' assessment of whether the transition is necessary and actionable.

## *Technology–Organisation–Environment and Protection Motivation Theory Lenses for Security-Technology Adoption*

This study combines the Technology–Organisation–Environment (TOE) framework with Protection Motivation Theory (PMT) to provide a comprehensive lens for analysing the factors influencing the adoption of ZTA in SMEs.

The TOE framework is particularly suited to understanding the organisational constraints and external pressures that impact technology adoption. The framework's three core dimensions—technological factors, organisational factors, and environmental factors—help explain why some SMEs may struggle to adopt ZTA while others are more inclined to adopt it. Technological complexity, organisational IT capability, and external environmental factors, such as regulatory pressure or supply chain security demands, can all influence the perceived benefits and feasibility of ZTA. For example, SMEs with limited technical resources may find the complexity of ZTA to be a significant barrier to adoption. Conversely, SMEs operating in highly regulated industries, such as healthcare or finance, may be more motivated to adopt ZTA to meet compliance requirements.

PMT, on the other hand, provides a psychological perspective by focusing on how decision-makers assess threats and coping strategies. Meta-analytic evidence in information security shows that coping appraisal variables—response efficacy and self-efficacy—exert the largest average effects on protective intentions, while perceived vulnerability and response cost are weaker or mixed (Mou et al., 2022; Sommestad et al., 2015). In the context of ZTA adoption, decision-makers in SMEs are likely to weigh the severity of potential cyber threats against the cost and complexity of implementing ZTA. PMT also highlights the importance of response efficacy (the belief that ZTA will mitigate security risks) and self-efficacy (the belief that the organisation has the capacity to implement ZTA). Understanding these psychological factors helps explain why some SMEs perceive ZTA as a necessary security measure while others may view it as an unattainable ideal due to perceived complexity or resource constraints.

The two lenses are combined because SME ZTA readiness is neither purely a technical implementation issue nor purely a psychological perception issue. TOE helps structure the technological, organisational, and environmental conditions that shape whether ZTA appears feasible for SMEs. PMT helps explain how decision-makers appraise security threats and the perceived effectiveness or actionability of ZTA as a response. In this study, TOE therefore captures feasibility conditions, while PMT captures threat and coping appraisals. Together, they provide a structured basis for examining perceived necessity and related ZTA perceptions in an exploratory pilot context.

**Table 1. Conceptual Mapping of TOE and PMT Lenses**

| | Analytical focus | Variables / themes in this study | Role in the analysis |
|---|---|---|---|
| TOE - Technology | Technical complexity and compatibility | IAM complexity, legacy integration, scalability concerns, cloud-computing need | Explains whether ZTA appears technically feasible and relevant |
| TOE - Organisation | Internal resources and capability | SME status, staff skills, barrier count, familiarity with ZTA | Explains organisational constraints and readiness conditions |
| TOE - Environment | External dependencies and pressures | Cloud dependence, regulatory and customer security expectations | Explains why ZTA may become salient for SMEs |
| PMT - Threat appraisal | Perceived exposure and seriousness of security threats | Threat experience, perceived threat severity | Explains why protective change may be considered |
| PMT - Coping appraisal | Perceived effectiveness and actionability of response | Response efficacy, familiarity, perceived necessity | Explains why ZTA may be viewed as useful or actionable |

# Methodology

## *Research Design*

This pilot study employs a dual-level comparative design to examine ZTA readiness.

Level 1 (Inter-group Benchmarking): The study includes large enterprises (LEs) as a comparative baseline. This design is not intended to propose a universal adoption path, but to scientifically identify the unique boundary conditions and resource-specific pressures faced by SMEs compared to their larger counterparts.

Level 2 (Intra-group Granularity): To address the diversity within the SME sector, the study further categorizes the SME sample into two cohorts to explore size-dependent variances in security readiness: Small Enterprises: 10–50 employees; Medium Enterprises: 50–250 employees.

## *Sample and Data Collection*

Participants were recruited using a purposive convenience sampling method (Ahmed, 2024). We administered an online questionnaire to information-technology and security professionals between 20 March and 5 May 2025. Recruitment used professional mailing lists and LinkedIn groups that target small-business IT administrators. Participation was voluntary and anonymous.

To ensure statistical rigor, an a priori power analysis was conducted using G*Power 3.1.9.7 to determine the required sample size. For a two-tailed t-test comparing two independent groups (SMEs vs. Large Enterprises) with a large effect size (d=0.8), a significance level (α) of 0.05, and a desired power level of 0.80, the minimum required total sample size was calculated as 52 (26 per group). Our final usable sample of N=64 (34 SMEs and 30 large enterprises) exceeds this statistical requirement, providing sufficient power to detect significant differences between the cohorts. Furthermore, as an exploratory pilot study, this sample size is consistent with similar information systems research targeting hard-to-reach professional populations.

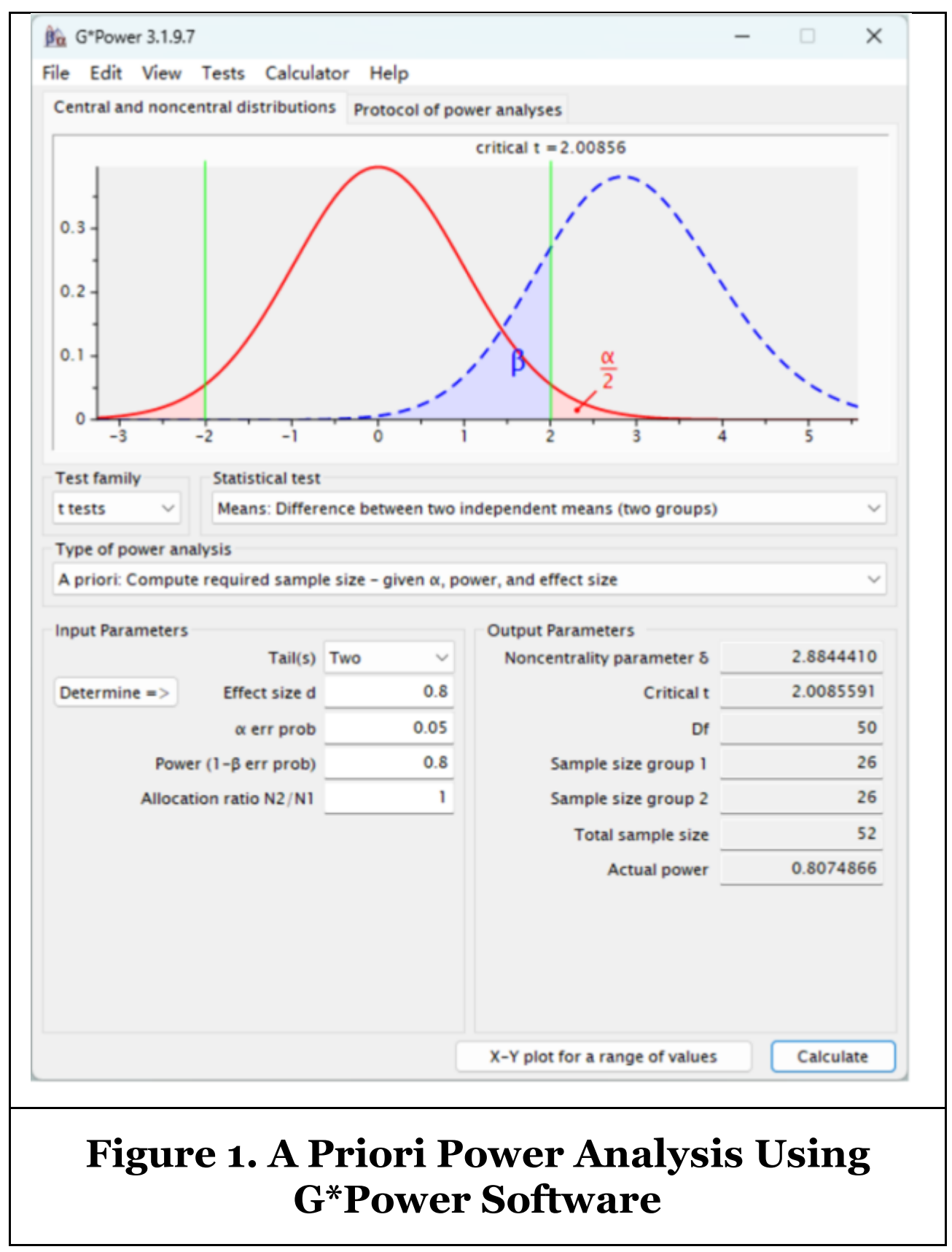

**Figure 1. A Priori Power Analysis Using G*Power Software**

A total of 64 usable cases remained after data cleaning, comprising 34 SMEs and 30 larger firms. While the total sample size meets the requirements for robust statistical modelling (e.g., OLS regression), the study is framed as an exploratory pilot study due to the inherent challenges in accessing specialized SME practitioners.

## *Questionnaire*

The questionnaire comprised six sections and 27 questions. The 27-item instrument was structured around four analytical themes derived from the TOE and PMT framing: contextual baseline, cognitive appraisals, readiness and advantage, and structural barriers.

Specifically, section 1 captures basic demographic and professional background variables, as these are proven to affect cybersecurity attitudes and practices (Turner & Ledwith, 2018). Section 2 focuses on current security practices and threat experience to establish a baseline for organisational context—this follows common practice in security posture studies. Section 3 targets perceptions and expected benefits of ZTA, in line with the "technology acceptance model" and frameworks measuring perceived value of IT innovations. Section 4 is dedicated to challenges, reflecting recurring findings in both academic and industry research that successful adoption of new security models depend on identifying and addressing key barriers (e.g., cost, complexity, skills). Section 5 compares ZTA to traditional models to assess whether respondents perceive a competitive advantage, echoing the comparative approach used in many cyber-risk and technology adoption studies. Section 6 invites open-ended feedback to ensure that unexpected or context-specific insights can be captured, which is important for qualitative validation and future instrument refinement.

Likert-type perceptual items employed five-point response scales ("Strongly disagree" = 1 to "Strongly agree" = 5), consistent with established social science methodology for capturing attitudes and perceptions (Koo & Yang, 2025; Likert, 1932).

The formulation of questions was initially based on the literature review: we hypothesize that SMEs may shy away from ZTA implementation due to implementation cost (Okta, 2023), integration difficulties, increased staff training and the need for cultural and structural challenges (Schneier & Vance, 2025); on the other hand, large enterprises have benefited from implementing ZTA. These benefits range from reduced potential widespread damage, reduction in data breaches and security incidents and future cost savings (Balaouras, n.d.; Mittal, 2023; Pulse Secure, 2020).

The questionnaire was developed through iterative consultation with academic supervisors and experienced practitioners to ensure clarity, relevance, and comprehensive coverage of both technical and organisational dimensions. Item wording and response options were adapted, where appropriate, from peer-reviewed instruments and theoretical models capturing ZTA familiarity, perceived necessity, satisfaction, perceived benefits, and perceived barriers, with modifications to fit the SME context (Jimmy, 2022; Schneier & Vance, 2025). The overall structure was informed by established research literature and international frameworks, notably NIST SP 800-207, and aligned with widely cited sectoral surveys (e.g., Forrester; Gartner) to facilitate statistical analysis and meaningful comparisons between SMEs and large enterprises while enabling benchmarking against recognised knowledge in the field.

## *Measures*

Table 2 summarises the constructs and variable coding. Perceptual constructs such as threat severity, response efficacy, response cost, familiarity with ZTA, perceived necessity, perceived competitive advantage, and satisfaction with the current security model were measured on five-point Likert scales (Likert, 1932). This method enables the capturing of human sentiments in a standardized manner (Koo & Yang, 2025). It also uses items adapted from Protection Motivation Theory and prior IS security adoption research, reviewed by three domain experts (two CISSP-certified consultants and one academic) for validity. Resource constraints and environmental pressures draw on ENISA SME survey items and questions on regulatory and governance demands. To keep the survey manageable for busy practitioners, several concepts are captured with single-item indicators, which limits psychometric validation and precision; others use simple counts or index variables (for example, the number of perceived barriers or recent incident types). All items are coded so that higher scores represent more of the labelled concept.

| **Table 2. Construct Measures and Variable Coding** | | |
|---|---|---|
| | Source Examples | Sample Items (coded) |
| Perceived Threat Severity | PMT–based scales (Boss et al., 2015) | “Cyber-attacks could severely disrupt my organisation’s operations” (Q2.1) |
| Coping (Response) Efficacy | PMT | “Implementing zero trust controls would effectively reduce breaches” (Q2.4) |
| Perceived Response Cost | PMT / TOE “Organisation” dimension | “The financial cost of ZTA is prohibitive for my organisation” (Q4.1) |
| Resource Constraints | ENISA SME survey items | “Our IT staff lack the skills to maintain ZTA solutions” (Q4.2) |
| Satisfaction with current security model | Extended technology-adoption scales | “We are satisfied with the current security model” (Q2.9_1) |

### *Data Analysis Procedures*

The dataset was pre-processed by removing the Qualtrics header row and converting Likert-scale strings to numerical values from 1 to 5. Responses to multi-select items (Q3.3 and Q4.1) were decomposed into binary indicator variables, and two counts were derived to capture perceived benefits and barriers. Categorical recoding produced an SME indicator defined as fewer than 250 employees, and security-model labels were harmonized to a consistent taxonomy. Missingness was examined and handled via listwise deletion in model-specific analyses.

Descriptive statistics summarized the sample composition, including size, industry, and tenure. Associations between security model and organisational size were assessed using chi-squared tests, with Fisher’s exact test applied when expected cell counts were small. Group differences between SMEs and non-SMEs on Likert outcomes (Q2.9_1, Q3.1_1, Q3.2_1, Q5.1_4) were tested using the Mann–Whitney U test. Monotonic relationships among Likert outcomes were evaluated with Spearman’s rank correlation coefficient ($\rho$). Selection rates for benefits and challenges were reported for the full sample and stratified by SME status.

Primary modelling treated five-point Likert outcomes as approximately continuous and estimated ordinary least squares regressions with heteroskedasticity-robust (HC1) standard errors. The dependent variable in regression analyses is the perceived necessity of ZTA (Q3.2_1). Predictors comprised ZTA familiarity (Q3.1_1), the barrier count, SME status, prior threat experience (Q2.5), cloud-computing need (Q2.1), and remote-access need (Q2.3). An exploratory interaction between familiarity and barrier count was included to probe whether familiarity conditioned the influence of perceived barriers. These analyses are exploratory, and results are interpreted cautiously.

Robustness was examined using proportional-odds ordered logistic regression with the same covariates and interaction term; adjacent response categories were merged where sparse to satisfy model assumptions. Results were concordant with the OLS estimates in sign and substantive interpretation. Variance inflation factors were examined for all predictors, indicating no material multicollinearity. All data preparation and analysis were conducted in Python using pandas, numpy, scipy, and statsmodels, with selected cross-checks replicated in SPSS.

## Results

### *Theme 1: Response Profile and Security Posture*

The refined dataset comprises 64 Information Technology and security professionals in the Asia-Pacific region, consisting of an SME cohort (N=34) and a Large Enterprise (LE) comparative baseline (N=30). To examine intra-group variance, the SME sample was further stratified into Small Enterprises (<50 employees, n=10) and Medium Enterprises (50–250 employees, n=24). Table 3 summarises the sample profile. The organisations span sectors such as technology, energy, entertainment and media, healthcare, manufacturing, education, hospitality, retail, construction, and transport and logistics. Respondents cover a broad range of IT and security experience levels, from early-career professionals to practitioners with more than 15 years in the field. As shown in Table 4, a cross-tabulation of security model in use by organisation size shows no statistically significant association ($\chi^2 = 4.27$, df = 6, p = .64).

**Table 3. Sample Profile (Panels A–C)**

| Panel A. Organisation Size | | |
|---|---|---|
| | N | % |
| Small (<50) | 10 | 15.6 |
| Medium (50–250) | 24 | 37.5 |
| Large (250–500) | 12 | 18.8 |
| Very large (>500) | 13 | 20.3 |
| N/A | 5 | 7.8 |
| Total | 64 | 100.0 |
| Panel B. Industry | | |
| | N | % |
| Technology | 12 | 18.8 |
| Energy | 12 | 18.8 |
| Entertainment and Media | 7 | 10.9 |
| Healthcare | 6 | 9.4 |
| Manufacturing | 5 | 7.8 |
| Education | 5 | 7.8 |
| Finance | 5 | 7.8 |
| Hospitality | 4 | 6.2 |
| Retail | 3 | 4.7 |
| Transportation and Logistics | 3 | 4.7 |
| Construction | 2 | 3.1 |
| Total | 64 | 100.0 |
| Panel C. Years of Experience | | |
| | N | % |
| 0–3 years | 15 | 23.4 |
| 4–7 years | 9 | 14.1 |
| 8–11 years | 12 | 18.8 |
| 12–15 years | 20 | 31.2 |
| > 15 years | 8 | 12.5 |
| Total | 64 | 100.0 |

| **Table 4. Security Model by Organisation Size (row %)** | | | | |
|---|---|---|---|---|
| | Perimeter | Zero Trust | Hybrid | N |
| Small (<50) | 0.300 | 0.500 | 0.200 | 10 |
| Medium (50–250) | 0.333 | 0.417 | 0.250 | 24 |
| Large (250–500) | 0.333 | 0.250 | 0.417 | 12 |
| Very large (>500) | 0.538 | 0.231 | 0.231 | 13 |
| N/A | 0.200 | 0.200 | 0.600 | 5 |

## *Theme 2: Readiness, Necessity, and Intra-SME Variance*

As shown in Table 5, respondents were generally neutral toward ZTA-related security change. Overall satisfaction with the current security model sits around neutral (M = 2.98), ZTA familiarity is modest (M = 3.02), perceived necessity is marginally above neutral (M = 3.06), and perceived competitive advantage is also close to neutral (M = 3.02). These levels indicate neither strong endorsement nor clear rejection of ZTA among participants. Given the pilot nature of this study (N = 64), non-parametric tests and medians were used to support cautious interpretation.

Comparing SMEs (<250 employees) with non-SMEs shows no statistically significant differences on any of the four outcomes (all p >= .45). Although the SME median for competitive advantage is 3 compared with 2 for non-SMEs, the difference is not statistically reliable. We therefore do not infer a group-level advantage for SMEs. Instead, the results suggest that respondents in this pilot sample remain in a cautious or neutral evaluation phase regarding ZTA and related security-model change.

| **Table 5. SME vs. Non-SME on Likert Outcomes (Mann–Whitney U, two-tailed, α = 0.05)** | | | | | | |
|---|---|---|---|---|---|---|
| | SME Mean (SD) | SME Median | Non-SME Mean (SD) | Non-SME Median | U | p |
| Satisfaction | 2.96 (0.91) | 3 | 2.91 (0.95) | 3 | 432.0 | 0.847 |
| ZTA familiarity | 3.08 (0.97) | 3 | 2.89 (0.90) | 3 | 465.5 | 0.457 |
| Perceived necessity | 3.04 (1.16) | 3 | 3.00 (0.97) | 3 | 428.0 | 0.899 |
| Competitive advantage | 3.08 (1.14) | 3 | 2.89 (0.99) | 2 | 460.5 | 0.495 |

To explore intra-SME variance, a Mann-Whitney U test was conducted to compare perceived necessity between Small and Medium cohorts. Results indicated no statistically significant difference between Small (Median = 4.0) and Medium enterprises (Median = 3.0) in perceived necessity for ZTA (U = 135.0, p = .554). Because the subgroup sizes are small (Small n = 10; Medium n = 24), this analysis should be interpreted as exploratory and descriptive. We therefore avoid inferring that strategic valuation is uniform across the SME sector. Rather, the pilot data provide no statistically reliable evidence of a difference between small and medium enterprises on perceived necessity.

Overall, these descriptive results motivate further exploratory analysis of associations among ZTA familiarity, perceived necessity, satisfaction with the current security model, and perceived competitive advantage, without implying causal mechanisms or validated adoption intentions.

## *Theme 3: Perceived Benefits and Structural Barriers*

To provide a rigorous assessment of the comparative baseline (Level 1 analysis), 95% Confidence Intervals (CIs) were calculated for the proportional data across both cohorts, as presented in Table 6.

The benchmarking highlights the boundary conditions of SMEs. While financial investment is a shared concern, SMEs more frequently reported operational friction related to Identity and Access Management (IAM) complexity (52.9% vs. 33.3%) and scalability issues (23.5% vs. 3.3%). Conversely, LEs more frequently reported staff training as a challenge (36.7% vs. 5.9% for SMEs), reflecting different socio-technical scaling concerns across the cohorts.

**Table 6. Comparative Analysis of Perceived Benefits and Challenges (Panels A–B)**

| Panel A. Benefits | | |
|---|---|---|
| | SME % [95% CI] | Non-SME % [95% CI] |
| Enhanced security | 52.9% [36.2%, 69.7%] | 33.3% [16.5%, 50.2%] |
| Reduced attack surface | 32.4% [16.6%, 48.1%] | 26.7% [10.8%, 42.5%] |
| Improved compliance | 23.5% [9.3%, 37.8%] | 13.3% [1.2%, 25.5%] |
| Minimized Lateral Movement | 17.6% [4.8%, 30.5%] | 26.7% [10.8%, 42.5%] |
| Panel B. Challenges | | |
| | SME % [95% CI] | Non-SME % [95% CI] |
| Complexity of IAM | 52.9% [36.2%, 69.7%] | 33.3% [16.5%, 50.2%] |
| Financial investment | 26.5% [11.6%, 41.3%] | 33.3% [16.5%, 50.2%] |
| Scalability issues | 23.5% [9.3%, 37.8%] | 3.3% [0.0%, 9.8%] |
| User experience concerns | 20.6% [7.0%, 34.2%] | 23.3% [8.2%, 38.5%] |
| Legacy system integration | 17.6% [4.8%, 30.5%] | 13.3% [1.2%, 25.5%] |
| Staff training | 5.9% [0.0%, 13.8%] | 36.7% [19.4%, 53.9%] |

### *Theme 4: Correlates of Perceived Necessity*

Table 7 reports Spearman correlations among the main Likert-scale outcomes. Correlations are large and positive among the core attitudinal variables: respondents who are more familiar with ZTA tend to rate it as more necessary, more competitively advantageous, and more satisfactory as a security-model direction relative to their current security model. These patterns suggest a coherent favourable orientation toward ZTA-related security change.

However, the high correlations also indicate that these variables may partly reflect a broader attitudinal orientation rather than fully distinct validated constructs. Therefore, the correlations are interpreted as exploratory associations. They do not establish causal relationships or psychometrically distinct latent constructs.

**Table 7. Spearman Correlations among Likert Outcomes**

| | ρ | p-value | N |
|---|---|---|---|
| Familiarity ↔ Necessity | 0.84 | < .001 | 61 |
| Familiarity ↔ Satisfaction | 0.93 | < .001 | 63 |
| Familiarity ↔ Competitive advantage | 0.85 | < .001 | 61 |
| Necessity ↔ Competitive advantage | 0.74 | < .001 | 61 |

To explore factors associated with perceived necessity, a heteroskedasticity-robust Ordinary Least Squares (OLS) regression was estimated in Table 8. Familiarity with ZTA principles and the organization's cloud-computing needs emerged as the strongest positive correlates of perceived necessity in this pilot sample. The cumulative count of perceived barriers exhibited a weak negative association that was not statistically significant. It is notable that these findings should be interpreted cautiously. They indicate exploratory associations with perceived necessity, not causal drivers of ZTA adoption. In particular, the strong role of familiarity may reflect both implementation literacy and a broader favourable orientation toward ZTA-related security change.

Robustness. Ordered-logit robustness checks produced qualitatively similar patterns to the OLS results.

**Table 8. Robust OLS Predicting Perceived Necessity**

| | Coef | SE (HC1) | t | p |
|---|---|---|---|---|
| Intercept | 0.465 | 0.267 | 1.75 | .081 |
| Familiarity | 0.918 | 0.077 | 11.87 | <.001 |
| Barrier count | −0.178 | 0.134 | −1.33 | .184 |
| SME (1 = yes) | −0.064 | 0.157 | −0.41 | .685 |
| Threat experience | −0.131 | 0.148 | −0.89 | .375 |
| Cloud need | 0.453 | 0.162 | 2.80 | .005 |
| Remote need | −0.148 | 0.160 | −0.93 | .353 |

Exploratory intra-SME analysis was also conducted to examine barrier accumulation. The barrier accumulation median for Small enterprises (Md = 2.0) was slightly higher than for Medium enterprises (Md = 1.0), although the difference was not statistically significant (U = 145.0, p = .284). Given the small subgroup sizes and the coarse nature of the aggregate barrier count, this finding should be interpreted descriptively. It suggests that future research should examine how specific barrier types, such as IAM complexity, financial cost, skill gaps, legacy integration, and scalability concerns, shape perceived necessity and implementation readiness.

# Discussion

## *Interpreting the Comparative Baseline*

Regarding RQ1, the dual-level comparative design provides an exploratory baseline for understanding how ZTA-related perceptions differ across organisational size categories. The results do not indicate statistically reliable differences between SMEs and non-SMEs on the main Likert outcomes. However, the descriptive patterns remain useful for identifying issues that may be salient for SMEs, particularly in relation to perceived implementation barriers.

The benefit and challenge selections suggest that SMEs and large enterprises may face different forms of implementation friction. For large enterprises, staff training appears more prominent, whereas SMEs more frequently identify Identity and Access Management (IAM) complexity and scalability concerns. These descriptive patterns should not be interpreted as definitive group differences, but they are consistent with the view that SMEs operate under different resource and capability conditions from larger organisations. In this sense, the SME context remains important for ZTA research because smaller firms often face similar external threat environments while having fewer internal resources for security architecture transition.

## *Intra-SME Variance and Perceived Necessity*

A second purpose of the pilot study was to explore whether small and medium enterprises differ in their perceived necessity of ZTA. The non-parametric analysis did not identify a statistically significant difference between Small and Medium enterprises (U = 135.0, p = .554). This indicates that the strategic valuation of Zero Trust is relatively uniform across the SME sector. This suggests that the perceived strategic relevance of ZTA may be relatively consistent across small and medium organisations, regardless of whether an enterprise has 10 or 200 employees.

However, given the small subgroup sizes, especially the Small enterprise group (n = 10), this result should be interpreted descriptively rather than as evidence of uniformity across the SME sector. Future studies with larger and more stratified SME samples are needed before stronger conclusions can be drawn about intra-SME variation.

### *Interpreting Correlates of Perceived Necessity*

Regarding RQ2, the positive association between ZTA familiarity and perceived necessity aligns with cumulative evidence from information-security applications of Protection Motivation Theory (PMT). Meta-analytic work shows that coping-appraisal constructs—response efficacy and self-efficacy—tend to exert larger and more reliable effects on protective intentions than threat-appraisal variables such as perceived vulnerability or response cost (Mou et al., 2022; Sommestad et al., 2015).

Interestingly, the accumulation of perceived barriers exhibited only a weak, non-significant negative association with perceived necessity (p=.184). The weak, negative association of barrier accumulation with necessity is consistent with a complexity lens, which predicts that unmanaged integration and operational load erode realized security benefits; conversely, complexity that is explicitly governed is less likely to suppress adoption (Schneier & Vance, 2025). These convergences support interpreting familiarity as a proxy for capability and implementation literacy, with complexity operating as a context-dependent friction rather than a categorical veto.

Environmental fit further clarifies the pattern of results. The salience of cloud-computing needs is consistent with the Technology-Organisation-Environment (TOE) framework's environment dimension, in which external dependencies and regulatory pressures shape adoption pay-offs. Contemporary policy instruments and sectoral guidance reinforce this logic by setting measurable ZTA objectives and timelines (DoD, 2022; OMB, 2023), while national reports detail persistent exposure and capability gaps among SMEs (ACSC, 2025; ENISA, 2024). In such settings, identity-, device-, and data-centric controls increase the marginal value of auditable governance and traceability, strengthening perceived necessity even where resources are constrained (NCSC, 2022; NZISM, 2023). This broader policy context is also consistent with our finding that cloud-related dependence may heighten the perceived necessity of auditable, identity-centric controls.

However, these associations should not be interpreted as causal mechanisms or as evidence of adoption intention. The high correlations among familiarity, perceived necessity, perceived competitive advantage, and satisfaction with the current security model suggest that the variables may partly reflect a broader favourable orientation toward ZTA-related security change. Accordingly, familiarity is best interpreted here as an exploratory correlate of perceived necessity, rather than as a validated mediator or causal driver.

### *An Initial Staged Adoption Framework for SMEs*

The pilot findings also inform a cautious practical synthesis. Because IAM complexity and scalability concerns appear salient among SME respondents, a wholesale ZTA transformation may be unrealistic for many resource-constrained organisations. Instead, the results support presenting a staged adoption framework that aligns ZTA guidance with SME capability constraints.

This framework is an initial framework that synthesises the pilot observations with existing ZTA guidance and prior research on SME cybersecurity constraints. The pilot data add value by indicating which issues appear particularly salient in the SME context: IAM complexity, scalability concerns, cloud-computing needs, and ZTA familiarity. These observations help translate general ZTA guidance into a more staged and governable sequence for SMEs.

As illustrated in Figure 2, Stage 1 focuses on identity governance as the initial anchor. This includes centralising identities, strengthening authentication, improving joiner-mover-leaver processes, and clarifying access policies for critical systems. Stage 2 focuses on segmented protection of high-value assets, prioritising systems and workflows where compromise would create disproportionate operational or regulatory harm. Stage 3 focuses on targeted monitoring and automation aligned with operational capacity, rather than attempting comprehensive continuous verification from the outset.

This sequencing is intended to help SMEs move toward ZTA principles incrementally while managing resource constraints. It is not a substitute for further empirical testing. Future research should validate and refine this framework using larger samples, longitudinal data, and operational security metrics.

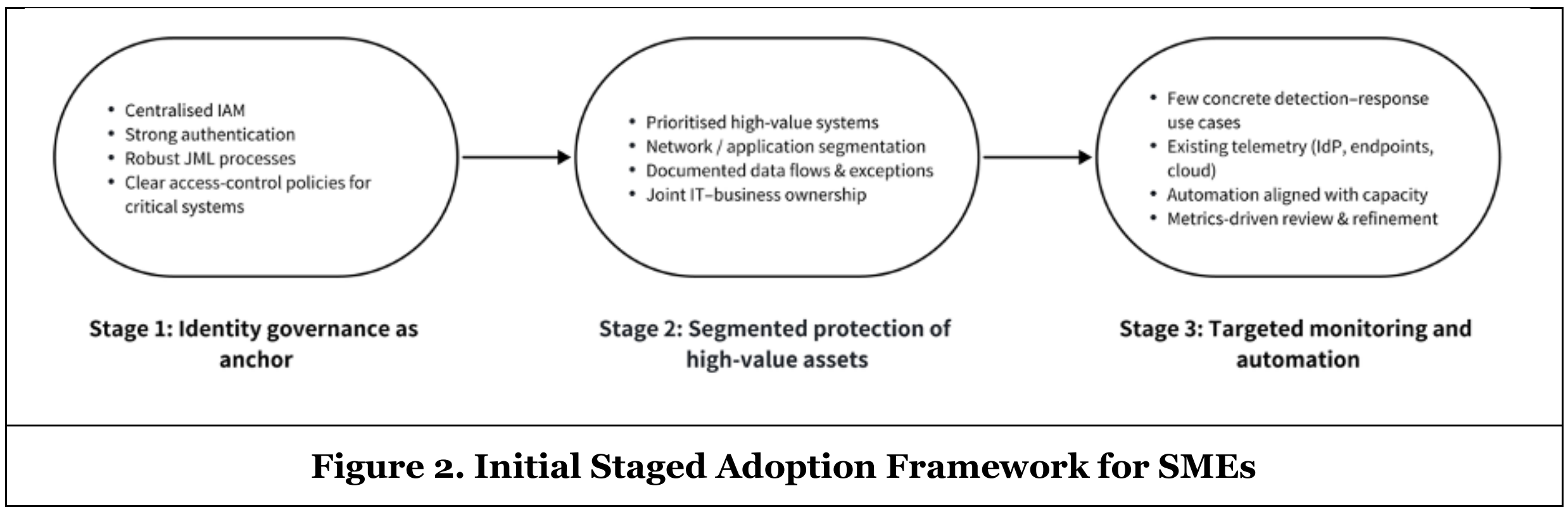

**Figure 2. Initial Staged Adoption Framework for SMEs**

# Implications

## *Theoretical Implications*

This study contributes cautiously to IS security adoption research by showing how TOE and PMT can be used together to structure an exploratory analysis of ZTA readiness among SMEs. TOE helps organise technological, organisational, and environmental conditions such as IAM complexity, resource constraints, cloud dependence, and scalability concerns. PMT helps interpret how decision-makers evaluate the necessity and actionability of ZTA under perceived threat and capability constraints.

This study highlights the value of integrating socio-technical adoption theories with concrete governance and complexity considerations when analysing emerging security architectures. First, familiarity appears as an exploratory correlate of perceived necessity, suggesting that future research could examine how implementation literacy shapes ZTA readiness formation. Second, the salience of cloud needs supports an environment–fit account of governance innovations, suggesting that ZTA yields higher marginal value where digital dependencies and service exposure are greater. Third, the weak barrier effect implies that complexity is not a categorical veto; rather, its influence is contingent on sequencing and scoping.

The findings do not establish new causal mechanisms. Instead, they suggest that familiarity with ZTA and cloud-computing needs are useful exploratory correlates of perceived necessity. This supports the value of studying ZTA readiness as both an organisational feasibility issue and a managerial appraisal issue. This study therefore positions SME's ZTA readiness as an exploratory socio-technical adoption problem rather than a purely technical implementation problem. SMEs may require adoption models that account for resource constraints, security capability, and staged governance, rather than assuming that large-enterprise ZTA blueprints can be transferred directly.

## *Practical Implications*

For practitioners and policymakers, the findings suggest that ZTA guidance for SMEs should emphasise staged and governable progress rather than uniform end-state architectures. SMEs may benefit from beginning with identity governance, because IAM complexity was one of the most frequently selected challenges among SME respondents. This does not mean that IAM is the only important control, but it provides a practical starting point that can support later segmentation and monitoring.

First, SMEs should consolidate identity and access management as the initial governance anchor. This includes centralising identities, enforcing strong authentication, improving joiner-mover-leaver processes, and clarifying role- or attribute-based access policies for business-critical systems. Assigning explicit ownership for identity data quality and access reviews can give SMEs a concrete focal point for both risk reduction and accountability.

Second, once identity governance is stabilised, SMEs can progressively introduce finer-grained protection for high-value systems. Rather than attempting to segment everything, organisations should prioritise a small number of critical systems and workflows where compromise would create disproportionate harm. This requires mapping data flows, documenting exceptions, and involving business stakeholders in approving and reviewing segmentation policies.

Third, SMEs should expand monitoring and automation in line with operational capacity. Continuous verification can begin with a limited number of concrete detection and response use cases, such as privileged account misuse or anomalous access from unmanaged devices. Existing telemetry from identity providers, endpoint tools, and cloud services can provide an initial basis for monitoring without requiring immediate enterprise-scale security operations capability.

Finally, external partners, vendors, and regulators can support SMEs by offering reference architectures and managed services that embody staged adoption. Guidance that links technical controls to organisational roles, accountability structures, and realistic resourcing levels is more likely to translate into sustainable Zero Trust practices for smaller organisations.

## Limitations

This study should be interpreted as an exploratory pilot study and has several limitations. First, although the final sample size (n = 64) is appropriate for the primary SME versus non-SME comparison, the intra-SME subgroup analysis is limited by the small number of Small enterprises (n = 10) and Medium enterprises (n = 24). These subgroup findings should therefore be read as descriptive rather than conclusive. Second, several focal variables, including perceived necessity, perceived competitive advantage, satisfaction with the current security model, and perceived barriers, were measured using single-item indicators or simple counts. This design helped keep the survey manageable for IT and security practitioners, but it limits construct precision and does not allow full reliability, convergent validity, or discriminant validity testing. The strong correlations among some attitudinal variables should therefore be interpreted cautiously. Third, the study relies on cross-sectional, self-reported survey data. As a result, the analysis identifies associations rather than causal relationships. In particular, perceived necessity is treated as the focal attitudinal outcome and should not be interpreted as a direct measure of adoption intention. Finally, the proposed staged adoption framework is an initial synthesis informed by the pilot findings and existing ZTA guidance. It is intended to support practical interpretation rather than to serve as a validated implementation model.

These constraints counsel caution in generalizing the findings and motivate future research. Future research should examine the framework using larger samples, validated multi-item measures, longitudinal designs, and operational security indicators.

## Conclusion

This pilot study examined how organisations, particularly SMEs, perceive and approach ZTA. Using a dual-level comparative design, it provided exploratory evidence on ZTA-related perceptions among SMEs and a large-enterprise comparison group, while also examining descriptive variation between small and medium enterprises.

The results suggest that ZTA familiarity and cloud-computing needs are positively associated with perceived necessity in this pilot sample. However, these findings should be interpreted as exploratory associations rather than causal drivers of adoption. The study also finds that SME respondents frequently identify IAM complexity and scalability concerns as implementation challenges, suggesting that ZTA guidance for SMEs may need to emphasise staged and governable progress rather than large-enterprise end-state blueprints.

The contribution of the study is therefore twofold. Theoretically, it shows how TOE and PMT can be used together to structure exploratory analysis of SME’s ZTA readiness, linking organisational feasibility conditions with decision-makers’ threat and coping appraisals. Practically, it offers an initial staged adoption framework that begins with identity governance, progresses to segmented protection of high-value assets, and then expands toward targeted monitoring aligned with operational capacity.

These contributions should be read in light of the pilot nature of the evidence. Future research should test the proposed framework with larger and more stratified samples, validated multi-item measures, longitudinal designs, and operational security metrics. Such work would help clarify whether the exploratory patterns identified here generalise across SME contexts and whether staged ZTA implementation improves measurable security outcomes over time.

## Acknowledgement

The authors gratefully acknowledge the IT and security professionals who participated in this survey. Their practical experience and perspectives provided valuable input for understanding Zero Trust Architecture readiness in SME contexts.